\documentclass[12pt]{article}

\newcommand{\be}{\begin{equation}}
\newcommand{\ee}{\end{equation}}
\newcommand{\bea}{\begin{eqnarray}}
\newcommand{\eea}{\end{eqnarray}}

\newcommand{\bit}{\begin{itemize}}
\newcommand{\eit}{\end{itemize}}

\newcommand{\beqs}{\begin{eqnarray}}
\newcommand{\eeqs}{\end{eqnarray}}
\begin{document}
\bibliographystyle{h-physrev}
{\sf \title{$W$ exchange contributions in hadronic decays of charmed baryons}
\author{Fayyazuddin\footnote{fayyazuddins@gmail.com}}
\maketitle
\begin{center}
\vspace{-1cm}
{\it Physics Department \\ Quaid-i-Azam University\\Islamabad 45320\\ Pakistan}
\end{center}

\begin{abstract}
The nonleptonic decays of  $\Xi_c^0$ and $\Lambda_c^+$ that are dominated by $W$-exchange are studied.  In particular, we anlayse the decay modes $\Xi_c^0  \rightarrow \Xi^{*-} \pi^+, \Sigma^{*+}K^-, \Omega^-K^+$ and $\Lambda_c^+\rightarrow \Sigma^0\pi^+, \Xi^0 K^+$, $\Xi^0_c\rightarrow \Sigma^+K^-$.
\end{abstract}
\vspace{-16cm}

\thispagestyle{empty}

\newpage

The nonleptonic decays of charmed baryons $B_c'(1/2 ^+):(\Lambda_c^+, \Xi_c^+, \Xi_c^0)$ belonging to the antisymmetric representation $\bar 3$ of $SU(3)$ are relevant, as only these baryons are stable against strong and electromagnetic interactions, with the exception of $\Omega_c^0$, one notes, since $B_c'(1/2 ^+)$ belong to the triplet representation $\bar 3$ of $SU(3)$ the matrix elements
\be
\langle B^*(3/2^+)|{\bar q}\gamma_\mu(1-\gamma_5)c|B_c'(1/2^+)\rangle= 0
\ee
where $q=u,d,s$.  Hence, factorization does not contribute to the decays $B_c'(1/2^+)\rightarrow B^*(3/2^+) + P$, thus the dominant contribution comes from $W$ exchange.

$W$ exchange in the nonrelativistic limit is encoded in the effective Hamiltonian \cite{FR, RF, FRbook}:
\be
H_W^{PC} = \frac{G_F}{\sqrt 2}V_{ud}V_{cs}\sum_{i\neq j}\alpha_i^{+}\gamma_j^{-}(1-\sigma_i\sigma_j)\delta^3(r), \label{NRH}
\ee
where $\alpha_i^+$ converts a $d$ quark into a $u$ quark and $\gamma_j^-$ converts a $c$ quark into an $s$ quark.  For the case under consideration:
\bea
\alpha_i|d\rangle &=& |u\rangle, \;\; i=1,2 \nonumber \\
\gamma_j |c\rangle &=& |s\rangle, \;\; j=3 \nonumber
\eea
 where $1,2$ specifies the position of the $d$ quark.  Note that $\Sigma_c^0, \Xi_c^{'0}$ belong to the symmetric $6$ representation of $SU(3)$ with spin wavefunction:
 \be
 \chi_{MS}^{1/2,1/2} =\frac{1}{\sqrt 6}|-(\uparrow\downarrow+\downarrow\uparrow)\uparrow + 2\uparrow\uparrow\downarrow\rangle
 \ee
 Whereas $\Lambda_c^+, \Xi_c^0$ belong to the antisymmetric representation $\bar 3$ of $SU(3)$ with spin wavefunction:
 \be
 \chi_{MA}^{1/2,1/2} =\frac{1}{\sqrt 2}|(\uparrow\downarrow-\downarrow\uparrow)\uparrow\rangle
 \ee
From this it follows that:
\bea
\left[\alpha_2^+\gamma_3^-(1-\sigma_2\cdot\sigma_3) + \alpha_1^+\gamma_3^-(1-\sigma_1\cdot\sigma_3)\right ]|\Lambda_c^+\rangle &=& \sqrt{2} |\Sigma^+\rangle \label{i}  \\
\left [\alpha_1^+\gamma_3^-(1-\sigma_1\cdot\sigma_3) + \alpha_2^+\gamma_3^-(1-\sigma_2\cdot\sigma_3)\right ]|\Xi_c^0\rangle &=& -\sqrt{2} |\Xi^0\rangle \label{ii} \\
\left[\alpha_2^+\gamma_3^-(1-\sigma_2\cdot\sigma_3) + \alpha_1^+\gamma_3^-(1-\sigma_1\cdot\sigma_3)\right ]|\Sigma_c^0\rangle &=& \sqrt{6} |\Sigma^0\rangle \label{iii}  \\
\left [\alpha_1^+\gamma_3^-(1-\sigma_1\cdot\sigma_3) + \alpha_2^+\gamma_3^-(1-\sigma_2\cdot\sigma_3)\right ]|\Xi_c^{'0}\rangle &=& -\sqrt{6} |\Xi^0\rangle \label{iv}
\eea

$W$ exchange is relevant when one considers the baryon-pole contributions (Born terms) involving the matrix elements of the form $\langle B|H_W^{PC}|B_c\rangle$, which can be evaluated in the nonrelativistic quark model (NQM) by using (\ref{NRH}).  The form of the Hamiltonian (\ref{NRH}) severely restricts which $B$ and $B_c$ lead to nonzero matrix elements.  The only nonzero matrix elements are:
\bea
\langle \Sigma^+|H_W^{PC}|\Lambda_c^+\rangle &=& (\frac{G_F}{\sqrt 2} V_{ud}V_{cs}){\sqrt 2}d' \label{2} \\
\langle \Xi^0|H_W^{PC}|\Xi_c^0\rangle &=& (\frac{G_F}{\sqrt 2} V_{ud}V_{cs}){\sqrt 2}d' \label{3}\\
\langle \Xi^0|H_W^{PC}|\Xi_c^{'0} \rangle &=& (\frac{G_F}{\sqrt 2} V_{ud}V_{cs}){\sqrt 6}d' \label{4}\\
\langle \Sigma^+|H_W^{PC}|\Sigma_c^{+} \rangle &=& \langle \Sigma^0|H_W^{PC}|\Sigma_c^{0} \rangle \\ &=& (\frac{G_F}{\sqrt 2} V_{ud}V_{cs}){\sqrt 6}d' \label{5}
\eea
where \cite{FR,DGG}
\be
d' = \frac{3(m_\Delta - m_N)}{8\pi \alpha_s}m_{cu}^2 \approx 5\times 10^{-3} GeV^{-3}. \label{6}
\ee
For the numerical value of $d'$, we have used the strong coupling constant $\alpha_s =0.5$, $m_u = 336 MeV, m_c = 1520 MeV, m_{cu} = \frac{m_u m_c}{m_c + m_u} \approx 275 MeV$.

For the decay of the type $B_c'(p')\rightarrow B^*(p) + P(q)$ \cite{FR}, the decay rate is given by \cite{FR}
\be
\Gamma = \frac{1}{6\pi}\frac{m'}{m^{*2}}\frac{|\bf p|^3}{f_p^2}(p_0+m^*)|C|^2
\ee
For  $\Lambda_c^+$ decays, the dominant contribution comes from the $\Sigma^+$ pole ({\it cf.} Eq (\ref{i}) ), i.e. from the chain
\be
\Lambda_c^+ \rightarrow \Sigma^+ \rightarrow \Delta^{++}K^-, \Sigma^{*0}\pi^+, \Xi^{*0}K^+
\ee
Hence the decay amplitude $C$ for these decays is given by (\ref{NRH}):
\be
C = (\sqrt 6, 1, \sqrt 2) g^*\frac{\langle \Sigma^+ |H_W^{PC}|\Lambda_c^+\rangle}{m_{\Lambda_c^+}-m_{\Sigma^+}}
\ee

Then from Equations (\ref{2},\ref{6}) and ${\sqrt 6}g^* \approx 2.09$:
\be
Br(\Lambda_c^+ \rightarrow \Delta^{++}K^-) \approx 1.04\times 10^{-2}
\ee
in good agreement with the experimental value \cite{PDG} ($1.09\pm 0.25)\times 10^{-2}$
\be
Br(\Lambda_c^+ \rightarrow \Sigma^{*0}\pi^+) \approx 2.2\times 10^{-3}
\ee
No experimental data is available to check the above prediction.  
\be
Br(\Lambda_c^+ \rightarrow \Xi^{*0}K^+) \approx 0.8\times 10^{-3}  \label{11}
\ee
is in disagreement with the experimental value \cite{PDG}:
\be
Br(\Lambda_c^+ \rightarrow \Xi^{*0}K^+) = (3.3\pm 0.9)\times 10^{-3}
\ee

The branching ratio in (\ref{11}) is in the $SU(3)$ limit.  Here we must take into account $SU(3)$ breaking compared to $\Lambda_c^+ \rightarrow \Delta^{++}K^-$, by multiplying the branching ratio given in  (\ref{11}) by a factor of $({m_s}/m_{cu})^2 \approx 3.4$.  This amounts to replacing $m_{cu}^2$ by $m_{cu}m_s$; $m_s=510MeV$.  With this modification
\be
Br(\Lambda_c^+ \rightarrow \Xi^{*0}K^+) \approx 2.7\times 10^{-3}
\ee
in agreement with the experimental value.

In this short note, the formulation developed in reference \cite{FR} is extended to $\Xi_c^0$.  For $\Xi_c^0$, the dominant contribution comes from the $\Xi^0$ pole({\it cf.} (\ref{ii})), i.e. from the chain:
\be
\Xi_c^0 \rightarrow \Xi^0 \rightarrow \Xi^{*-}\pi^+, \Sigma^{*+}K^-, \Omega^-K^+
\ee
Hence the decay amplitude $C$ for these decays is given by
\be
C = (\sqrt 2, -\sqrt 2, -\sqrt 6) g^*\frac{\langle \Xi^0 |H_W^{PC}|\Xi_c^0\rangle}{m_{\Xi_c^0}-m_{\Xi^0}} \label{13}
\ee
From (\ref{3}), using the value of $d'$ given in (\ref{6}), one gets:
\bea
\Gamma (\Xi_c^0 \rightarrow \Xi^{*-}\pi^+) &=& 2.2\times10^{-14}GeV \nonumber \\ 
\Gamma (\Xi_c^0 \rightarrow \Sigma^{*+}K^-) &=& 1.6\times10^{-14}GeV \\
\Gamma (\Xi_c^0 \rightarrow \Omega^-K^+) &=& 1.3\times10^{-14}GeV \nonumber
\eea
From the experimental value $\tau_{\Xi_c^0}\approx112\times 10^{-15}s, \Gamma_{\Xi_c^0}\approx 5.9\times 10^{-12} GeV$. Hence one gets the branching ratios:
\bea
Br (\Xi_c^0 \rightarrow \Xi^{*-}\pi^+) &=& 3.8\times10^{-3} \nonumber \\ 
Br (\Xi_c^0 \rightarrow \Sigma^{*+}K^-) &=& 2.7\times10^{-3}\\
Br (\Xi_c^0 \rightarrow \Omega^-K^+) &=& 2.3\times10^{-3} \nonumber
\eea
However, for $\Xi_c^0$, $m_{cu}^2$ is replaced by $m_{cu}m_{cs}$, i.e. by a factor
\be
\frac{m_{cs}}{m_{cu}} = \frac{0.382}{0.275} \approx 1.4
\ee
Thus the branching ratios are given by
\bea
Br (\Xi_c^0 \rightarrow \Xi^{*-}\pi^+) &=& 5.3\times10^{-3} \nonumber \\ 
Br (\Xi_c^0 \rightarrow \Sigma^{*+}K^-) &=& 3.8\times10^{-3}\\
Br (\Xi_c^0 \rightarrow \Omega^-K^+) &=& 6.3\times10^{-3} \nonumber
\eea
Currently, there is no experimental data available to check the above branching ratios.

Another aspect of this short note is to analyse the nonleptonic decays $B_c'(1/2^+)\rightarrow B(1/2^+) + P$ in which factorization does not contribute.  For such decays, the baryon pole contribution to the $p$-wave decay (parity conserving) amplitude $B$ is the dominant contribution.  

For the decays
\be
\Lambda_c^+ \rightarrow \Sigma^0\pi^+, \Xi^0K^+, \;\; p'=p+k
\ee
the dominant contribution comes from the baryon pole.  Hence the decay amplitude for $\Lambda_c^+ \rightarrow \Sigma^0\pi^+$:
\be
B = B(pole) = g_{\Lambda_c^+\pi^+\Sigma_c^0}\frac{1}{m_{\Sigma_c^0}-m_{\Sigma^0}}\langle \Sigma^0 |H_W^{PC}|\Sigma_c^0\rangle \label{18}
\ee
and for $\Lambda_c^+ \rightarrow \Xi^0K^+$:
\be
B = B(pole) = g_{\Lambda_c^+K^+\Xi_c^{'0}}\frac{1}{m_{\Xi_c^{'0}}-m_{\Xi^0}}\langle \Xi^0 |H_W^{PC}|\Xi_c^{'0}\rangle \label{19}
\ee
The Goldberger-Trieman relation gives
\bea
g_{\Lambda_c^+\pi^+\Sigma_c^{0}} &=& \frac{m_{\Lambda_c^+}+m_{\Sigma_c^0}}{F_\pi}g_A \label{20} \\
g_{\Lambda_c^+K^+\Xi_c^{'0}} &=& \frac{m_{\Lambda_c^+}+m_{\Xi_c^{'0}}}{F_K}g'_A \label{21} 
\eea
The quark model gives:
\be
g_A=-{\sqrt\frac{2}{3}},\;\; g'_A=-{\sqrt\frac{1}{3}} \label{22}
\ee
Hence from equations (\ref{5},\ref{18},\ref{20},\ref{22})
\be
B = B(pole) = \frac{G_F}{\sqrt 2}\frac{m_{\Lambda_c^+}+m_{\Sigma_c^0}}{m_{\Sigma_c^0}-m_{\Sigma^0}}\frac{2d'}{F_\pi} \label{23}
\ee
for the decay $\Lambda_c^+\rightarrow\Sigma^0\pi^+$.
For the decay $\Lambda_c^+\rightarrow \Xi^0K^+$, from equations (\ref{4},\ref{19}, \ref{21}, \ref{22})
\be
B = B(pole) = \frac{G_F}{\sqrt 2}\frac{m_{\Lambda_c^+}+m_{\Xi_c^{'0}}}{m_{\Xi_c^{'0}}-m_{\Xi^0}}\frac{2d'}{F_K}  \label{24}
\ee
From eq. (\ref{23}), using the value of $d'$ given in eq. (\ref{6}), one gets
\be
B^2(pole) \approx 5.58\times 10^{-12}, {\mbox for} \Lambda_c^+ \rightarrow\Sigma^0\pi^+ \label{25}
\ee
and from equation (\ref{24})
\be
B^2(pole) \approx 1.93\times 10^{-12}, {\mbox for} \Lambda_c^+ \rightarrow\Xi^0 K^+ \label{26}
\ee

For the decay \cite{FRbook}
\be
B_c'(p')\rightarrow B(p) + P(k)
\ee
the parity-conserving (p-wave) decay rate is given by
\be
\Gamma = \frac{k}{4\pi m'}(p_0-m)|B|^2. \label{27}
\ee
Hence from equations (\ref{25}) and (\ref{27}), the branching ratio
\be
Br(\Lambda_c^+ \rightarrow\Sigma^0\pi^+) \approx 1.26\% \label{28}
\ee
in agreement with the experimental value $(1.29\pm 0.47)\%$.  

From equations (\ref{26}) and (\ref{27}), the branching ratio
\be
Br(\Lambda_c^+ \rightarrow\Xi^0 K^+) \approx 2.0\times 10^{-3} \label{29}
\ee
to be compared with the experimental value: $(5.0\pm 1.2)\times 10^{-3}$.
The branching ratio in eq. (\ref{29}) is in the $SU(3)$ limit.  To take into account $SU(3)$ breaking, the branching ratio in eq. (\ref{29}) is multiplied by a factor $(m_{cs}/m_{cu})^2\approx 2$.  With this modification
\be
Br(\Lambda_c^+ \rightarrow\Xi^0 K^+) \approx 4.0\times 10^{-3} \label{30}
\ee
in agreement with the experimental value.

Finally, for the decay $\Xi_c^0\rightarrow \Sigma^+ K^-$, the dominant contribution comes from the baryon pole:
\be
B = B(pole) = g_{\Xi_c^0K^-\Sigma_c^+}\frac{1}{m_{\Xi_c^0}-m_{\Sigma^+}} 
\langle\Sigma^+|H_W^{PC}|\Sigma_c^+\rangle \label{31}
\ee
with 
\be
g_{\Xi_c^0K^-\Sigma_c^+} = \frac{m_{\Xi_c^0}+m_{\Sigma_c^+}}{F_K}g_A, \;\; g_{A}=\frac{1}{\sqrt 3}.
\label{32}
\ee
Hence from equations (\ref{5}), (\ref{31}) and (\ref{32})
\bea
B &=& B(pole) = \frac{G_F}{\sqrt 2}\frac{m_{\Xi_c^0}+m_{\Sigma_c^+}}{m_{\Xi_c^0}-m_{\Sigma^+}} \frac{1}{F_K}({\sqrt 2} d') \label{33} \\
B^2(pole) &\approx& 1.74\times 10^{-12} \nonumber
\eea
Equation (\ref{27}) gives
\be
\Gamma\approx 1.74\times 10^{-14} GeV \nonumber
\ee
\bea
Br(\Xi_c^0 \rightarrow\Sigma^+ K^-) &\approx & 2.9\times 10^{-3} \nonumber \\
&\rightarrow &  2.9\times 10^{-3} \times 1.4 \approx 4.1\times 10^{-3} \nonumber \\
&=& Br(\Xi_c^0 \rightarrow\Sigma^0 {\bar K}^0) = Br(\Xi_c^0 \rightarrow\Sigma^0 K_S) \label{34}
\eea

To conclude, in this note the formalism developed in \cite{FR} is extended to study the nonleptonic decays
\be
\Lambda_c^+ \rightarrow \Sigma^0 K^+,\Xi^0 K^+
\ee
for which the baryon poles contribute for the parity-conserving (p-wave) amplitude $B$.  The results obtained are in good agreement with the experimentally determined values. 
The same formalism is also used to analyse the decays 
\be
\Xi_c^0 \rightarrow \Xi^{*-} \pi^+, \Sigma^{*+} K^-, \Omega^{-} K^+.
\ee
and 
\be
\Xi_c^0 \rightarrow \Sigma^{+} K^-.  
\ee
For the $\Xi_c^0$ decays, experimental data is not available to check our results.  These results may be of interest to experimental physicists investigating charmed baryon decays.  The formalism developed in \cite{FR} based on \cite{RF}, in which the effective Hamiltonian given in (\ref{NRH}) plays a crucial role to analyse the decays for which $W$ exchange gives the dominant contribution and factorization has no role seems to be well established.

\end{document}